# Rubidium-Rich Asymptotic Giant Branch Stars

D. A. García-Hernández[1,*,†], P. García-Lario[1,2] B. Plez[3], F. D'Antona[4], A. Manchado[5],

and J. M. Trigo-Rodríguez[6,7]

[1] ISO Data Centre. European Space Astronomy Centre, Research and Scientific Support Department of ESA. Villafranca del Castillo. Apdo. 50727. E-28080 Madrid, Spain

[2] Herschel Science Centre. European Space Astronomy Centre, Research and Scientific Support Department of ESA. Villafranca del Castillo. Apdo. 50727. E-28080 Madrid, Spain

[3] GRAAL, UMR 5024, Université de Montpellier 2, F-34095 Montpellier Cedex 5, France

[4] INAF- Osservatorio Astronomico di Roma, via Frascati 33, I-00040 MontePorzio Catone, Italy

[5] Instituto de Astrofísica de Canarias, La Laguna, E-38200, Tenerife, Spain and Consejo Superior de Investigaciones Científicas

[6] Institute of Space Sciences (CSIC), Campus UAB, Facultat de Ciències, Torre C-5, parells, 2ª planta, 08193 Bellaterra, Barcelona, Spain

[7] Institut d'Estudis Espacials de Catalunya (IEEC), Ed. Nexus, Gran Capità 2-4, 08034 Barcelona, Spain

* To whom correspondence should be addressed. E-mail: agarcia@astro.as.utexas.edu

† Present address: The W. J. McDonald Observatory. The University of Texas at Austin. 1 University Station, C1400. Austin, TX 78712-0259, USA

A long debated issue concerning the nucleosynthesis of neutron-rich elements in Asymptotic Giant Branch (AGB) stars is the identification of the neutron source. We report intermediate-mass (4 to 8 solar masses) AGB stars in our Galaxy that are rubidium-rich owing to overproduction of the long-lived radioactive isotope $^{87}$Rb, as predicted theoretically 40 years ago. This represents a direct observational evidence that the $^{22}$Ne($\alpha$,n)$^{25}$Mg reaction must be the dominant neutron source in these stars. These stars then challenge our understanding of the late stages of the evolution of intermediate-mass stars and would promote a highly variable Rb/Sr environment in the early solar nebula.

Low- and intermediate-mass (1−8 solar mass, $M_\odot$) stars evolve towards the Asymptotic Giant Branch (AGB) phase (1) after the completion of hydrogen and helium burning in their cores, before they form Planetary Nebulae, ending their lives as White Dwarfs. Basically, an AGB star is composed of an inert carbon-oxygen (C-O) core surrounded by a He-rich intershell and an extended H-rich convective envelope. Nuclear energy release is dominated by the H-shell and interrupted periodically by thermonuclear runaway He-shell "thermal pulses", that initiate a series of convective and other mixing events. Strong mass loss enriches the interstellar medium (ISM) with the products of the complex resulting nucleosynthesis (2). During this thermally pulsing AGB (TP-AGB) phase, stars originally born O-rich reflecting the ISM composition, can turn C-rich (C/O > 1) as a consequence of the "dredge-up" of processed material from the bottom of the convective envelope to the stellar surface. In AGB stars of higher/intermediate mass (4−8 $M_\odot$) the convective envelope penetrates the H-burning shell activating the so-called "hot bottom burning" (HBB, hereafter) process (3, 4). HBB takes place when the temperature at the base of the convective envelope is hot enough (T $\geq$ 2 x $10^7$ K) that $^{12}$C can be converted into $^{13}$C and $^{14}$N through the CN cycle, so these AGBs are no longer C-stars and become again or remain O-rich in spite of the dredge-up. HBB models (3, 4) predict also the production of the short-lived $^7$Li isotope, through the "$^7$Be transport mechanism" (5), which should be detectable at the stellar surface. The HBB activation in massive AGB stars is supported by lithium overabundances in luminous O-rich AGB stars of the Magellanic Clouds (6, 7). In our own Galaxy, a small group of stars showing OH maser emission at 1612 MHz (sometimes without optical counterpart but very bright in the infrared, the OH/IR stars) has recently been found to show strong Li abundances (8).

Mixing of protons into the He-rich intershell during the TP-AGB phase leads to reaction chains producing free neutrons, which allow production of neutron-rich elements like Rb, Sr, Y, Zr, Ba, La, Nd, Tc, etc. by slow-neutron captures on iron nuclei and other heavy elements (the s-process) (9-11). There are two possible chains for the neutron production: $^{13}$C($\alpha$,n)$^{16}$O and $^{22}$Ne($\alpha$,n)$^{25}$Mg. The $^{13}$C neutron source operates at relatively low neutron densities ($N_n < 10^7$ cm$^{-3}$) and temperatures T < 0.9-1 x $10^8$ K (11, 12) in TP-AGB stars during the interpulse period, the $^{22}$Ne neutron source operates at much higher neutron densities ($N_n > 10^{10}$ cm$^{-3}$) and requires higher temperatures (T > 3.0 x $10^8$ K) which are only achieved while the convective thermal pulse is ongoing. In the more massive AGB stars (> 4−5 M$_\odot$), where these high temperatures are more easily achieved, the s-process elements are expected to form mainly through the $^{22}$Ne($\alpha$,n)$^{25}$Mg reaction (11, 13). The $^{22}$Ne neutron source strongly favours as well the production of the stable isotope [87]Rb because of the operation of a branching in the s-process path at [85]Kr (14) which modifies the isotopic mix between [85]Rb and [87]Rb (14-17). Unfortunately, the Rb isotope ratio cannot be measured in stellar sources (17) even with the help of very high-resolution spectra because the lines are too broad, and, thus, it is difficult to use this parameter as a neutron density indicator. As an alternative, the total Rb abundance can be used. The theoretical prediction is that the relative abundance of Rb to other nearby s-elements such as Zr, Y and Sr, is a powerful indicator of the neutron density at the s-process site and, as such, a good discriminant of the operation of the [13]C versus the [22]Ne neutron source in AGB stars (11, 15-17).

Our sample is composed of 102 Galactic OH/IR stars, recently identified by us as massive O-rich AGB stars, for which we recently determined their Li and Zr abundances (8). These stars are experiencing very strong mass loss rates (up to several times $10^{-5}$ $M_{\odot}yr^{-1}$) at this stage and, as a consequence of this, most of them are heavily obscured by thick circumstellar envelopes, making optical observations very difficult. Indeed, 42 stars were found to be too faint at 7800 Å to perform any kind of analysis.

Despite this observational problem, we were able to obtain high-resolution optical echelle spectra (resolving power of ~40,000−50,000) for 60 stars in the sample. The observations were carried out using the Utrecht Echelle Spectrograph at the 4.2m William Herschel Telescope at the Observatorio del Roque de los Muchachos (La Palma, Spain) during three different observing runs in August 1996, June 1997 and August 1997 and the CAssegrain Echelle SPECtrograph of the European Sourthern Observatory 3.6m telescope at La Silla, Chile in February 1997. Because of the very red colours of the sources observed, the Signal-to-Noise (S/N) ratios achieved in the reduced spectra can strongly vary from the blue to the red orders (10−20 at 6000 Å while >100 at 8000 Å). The AGB stars studied here are quite different from other galactic AGB samples previously studied because they show the coolest temperatures yet observed ($T_{eff}$~2700−3300 K) and display peculiar properties. The extremely red spectra are dominated by strong molecular bands mainly due to titanium oxide (TiO). The TiO veiling effect is so intense that it is very difficult to identify individual atomic lines in the spectra of these stars with the exception of neutral species like e.g.: Li I (6708 Å), Ca I (6122 and 6573 Å), K I (7699 Å), Rb I (7800 Å) and a few Fe I lines. From the measurement of the radial velocities, we conclude that the Li and Ca atomic lines, as well as the TiO molecular bands, must be formed in the stellar atmosphere, while the K I and Rb I absorption lines usually have circumstellar components with peculiar velocities. For the majority of stars, the difference between the mean radial velocities of the stellar and circumstellar lines is of the order of the expansion velocity of the circumstellar envelope, as derived from the OH maser measurements. In particular, the Rb I resonance line at 7800.3 Å is originated not only in the photosphere of these stars but also in the outer non-static layers of the stellar atmosphere, and in the expanding circumstellar shell, as our observations confirm.

By using state-of-the-art synthetic models appropiate for cool O-rich AGB spectra we first derived the stellar fundamental parameters (e.g. effective temperature $T_{eff}$~2700−3300 K, solar metallicity [Fe/H]=0.0, C/O=0.5, gravity logg=−0.5, etc.) of these stars (8). Then we carried out a chemical abundance analysis on those sources for which a spectrum with enough S/N was obtained in the 7775-7835 Å region around the Rb I 7800.3 Å line. For this we constructed a grid of model spectra at different effective temperatures in the 7775-7835 Å region and we determined by χ-squared minimisation which of the model spectra provided the best fit to the observations. The goal was to fit the overall shape of the spectra including the TiO bandheads, which are very sensitive to variations in the effective temperature as well as the Rb abundances, which were derived by fitting the Rb I line. The Rb hyperfine structure (18) was considered, for which we assumed a solar Rb isotopic composition (19). Note that an extremely different value of the Rb isotope ratio only changes the derived Rb abundances by a maximum of 0.1 dex. A few sample spectra together with the best model fits are presented in Fig.1. We could obtain reliable Rb photospheric abundances for 22 stars in our sample (Table 1). The rest of stars displayed also very strong Rb I lines, but the circumstellar contribution to the line was difficult to quantify and thus an accurate abundance analysis was not possible. Our analysis yielded a wide variety of Rb abundances ranging from [Rb/Fe]~ −1.0 to +2.6 dex. The overall uncertainty in the derived Rb abundances is estimated to be always less than 0.8 dex. This mainly reflects the sensitivity to changes in the atmospheric parameters adopted for the modelling.

**Table 1. Spectroscopic effective temperatures and Rb abundances derived**. When the Rb I line at 7800.3 Å is resolved in two components (circumstellar and stellar), the abundance estimate corresponds to the photospheric abundance needed to fit the stellar component and it is marked with an asterisk. Stars IRAS 19059-2219 and IRAS 19426+4342 were observed in two different epochs. Note that the stars IRAS 10261-5055 and IRAS 19147+5004 are not long period, Mira-like variables and, as such, they are possibly non-AGB stars.

| IRAS name | Other | $T_{eff}$ (K) | [Rb/Fe] | S/N at 7800 Å |
|-----------|-------|---------------|---------|----------------|
| 01085+3022 | AW Psc | 3000 | +2.0 | 49 |
| 04404-7427 | SY Men | 3000 | +1.3* | 68 |
| 05027-2158 | T Lep | 2800 | +0.4 | 418 |
| 05098-6422 | U Dor | 3000 | +0.1 | 309 |
| 05151+6312 | BW Cam | 3000 | +2.1 | 161 |
| 06300+6058 | Ap Lyn | 3000 | +1.6* | 127 |
| 07222-2005 | ... | 3000 | +0.6* | 30 |
| 09194-4518 | MQ Vel | 3000 | +1.1* | 25 |
| 10261-5055 | VZ Vel | 3000 | <-1.0 | 595 |
| 14266-4211 | ... | 2900 | +0.9 | 106 |
| 15193+3132 | V S Crb | 2800 | -0.3 | 266 |
| 15576-1212 | Fs Lib | 3000 | +1.5 | 91 |
| 16030-5156 | V352 Nor | 3000 | +1.3 | 86 |
| 16037+4218 | V1012 Her | 2900 | +0.6 | 115 |
| 17034-1024 | V850 Oph | 3300 | +0.2 | 189 |
| 18429-1721 | V3952Sgr | 3000 | +1.2 | 98 |
| 19059-2219 | V3880 Sgr | 3000 | +2.3/+2.6 | 32/49 |
| 19147+5004 | TZ Cyg | 3000 | -0.5 | 110 |
| 19426+4342 | ... | 3000 | +1.0*/+1.0* | 19/25 |
| 20052+0554 | V1416 Aql | 3000 | +1.5* | 47 |
| 20077-0625 | V1300 Aql | 3000 | +1.3* | 19 |
| 20343-3020 | RT Mic | 3000 | +0.9 | 76 |

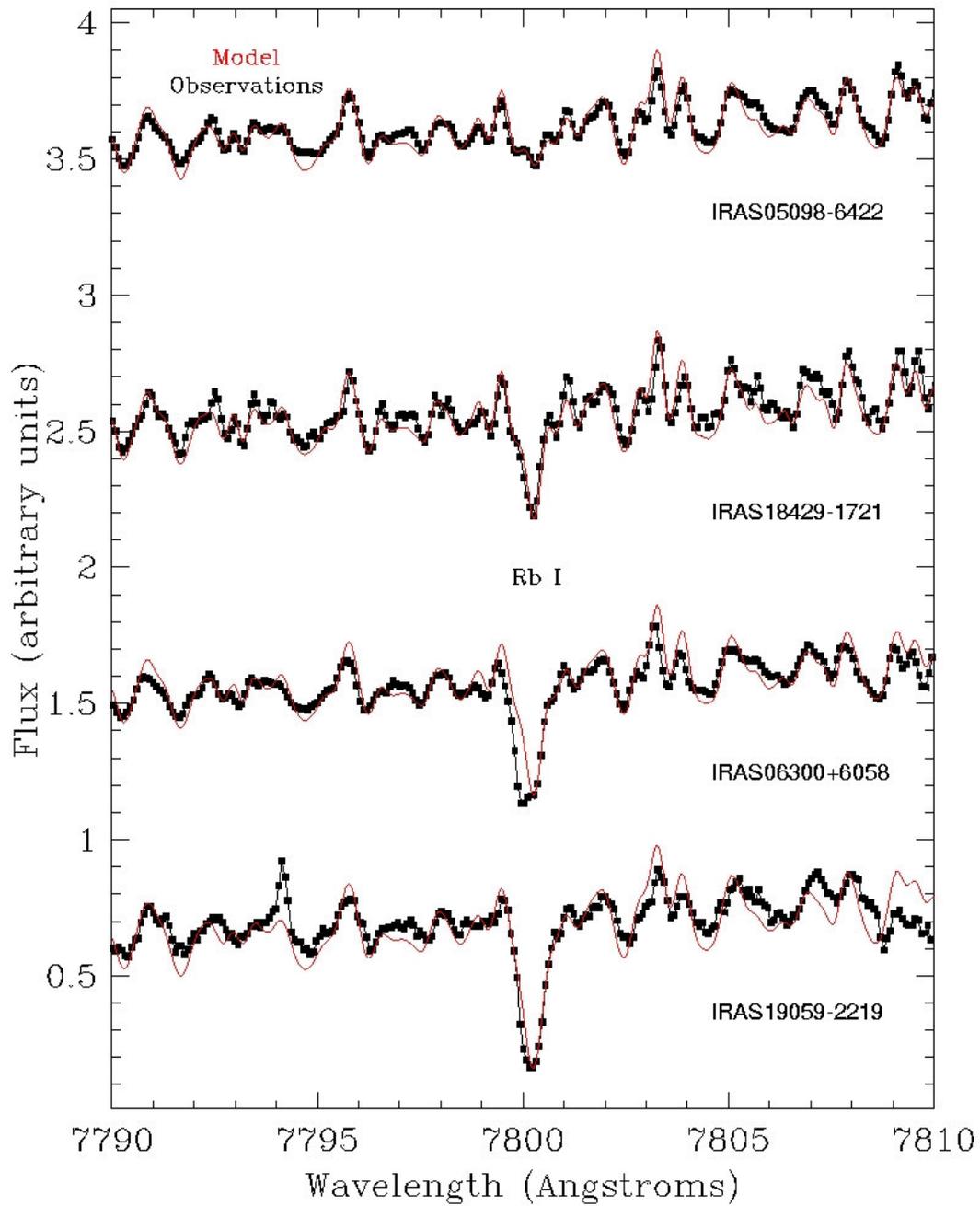

**Fig. 1. Best model fit and observed spectra around the Rb I line at 7800.3 Å.** Here the observed spectra (in black) and the best model fit (in red) of four sample stars (IRAS 05098-6422,

IRAS 18429-1721, IRAS 06300+6058 and IRAS 19059-2219) with an effective temperature of

3000 K but with a very different Rb abundance of [Rb/Fe]=+0.1, +1.2, +1.6 and +2.3 dex, respectively, are shown. Note that the Rb I resonance line has a clear "blue-shifted" circumstellar component in the Rb-rich star IRAS 06300+6058.

Using theoretical models the mass of the stars can be inferred from the observed nucleosynthesis pattern (15-17). Values of [Rb/Fe]≈−0.3−+0.6 dex and [Rb/Zr]<0, generally found in MS, S, and C (N-type) AGB stars in our Galaxy, are usually interpreted as an indication of these stars being low-mass (~1−3 $M_\odot$) AGB stars and of the $^{13}$C($\alpha$,n)$^{16}$O reaction being the main neutron source at the origin of s-process nucleosynthesis (15, 16). Instead, most of the stars analysed in our sample show a strong Rb enhancement ([Rb/Fe]~+0.6−+2.6 dex) in combination with only a mild Zr enrichment, as we determined [Zr/Fe] < 0.5 in our stars (8). The high Rb/Zr ratios (~0.1−2.1 dex) derived confirm that the stars in our sample belong to the group of more massive (> 4−5 $M_\odot$) AGB stars in our Galaxy and provides observational evidence that the $^{22}$Ne($\alpha$,n)$^{25}$Mg reaction is indeed the dominant neutron source in these AGB stars.

The expansion velocities derived from the OH masers can also be taken as an additional distance-independent mass indicator (20, 21). Despite the relatively large uncertainties involved, the Rb abundances were found to show a very nice correlation with the OH expansion velocities (Fig.2). This correlation seems to confirm that the efficiency of the $^{22}$Ne neutron source is directly correlated with the stellar mass, as a consequence of the higher temperature achieved in the He intershell during the convective thermal pulses, as predicted by the models (11-13). In Fig.2 we can clearly distinguish two groups of stars. Those stars in the sample with $v_{exp}$(OH) below 6 km s$^{-1}$ were identified as non-HBB AGBs (< 4−5 M$_\odot$) displaying relatively low Rb enhancements; while those with $v_{exp}$(OH) larger than 6 km s$^{-1}$, are suggested to be more massive AGB stars (> 4−5 M$_\odot$) experiencing HBB (8). Indeed, the more extreme stars in our sample, showing the larger Rb enhancements, must represent a class of even higher mass stars (~6-8 M$_\odot$). In particular, the extremely Rb-rich star IRAS 19059−2219 is probably the most massive AGB star in our sample, with Rb/Fe = +2.6 dex.

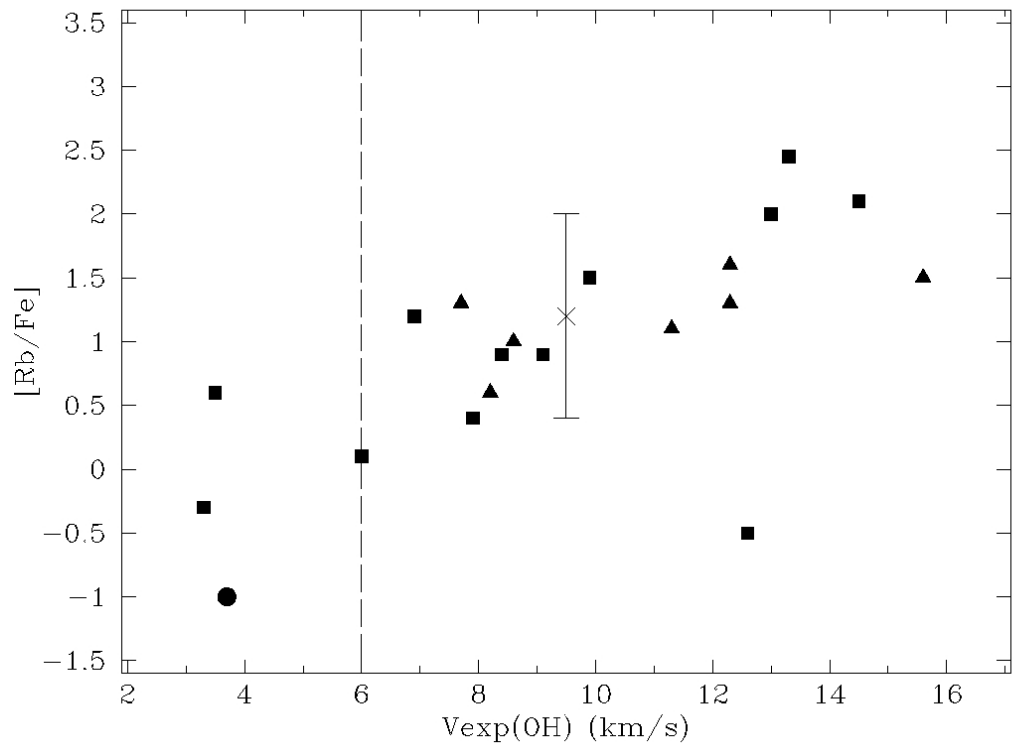

**Fig. 2. Observed Rb abundances versus OH expansion velocity.** One upper limit to the Rb abundance is shown with a dot . The abundance estimates that correspond to the photospheric abundance needed to fit the stellar component are shown with triangles. A maximum error bar of ±0.8 dex is also shown for comparison. The star with high OH expansion velocity and no Rb

which does not follow the general behaviour observed is IRAS 19147+5004 (TZ Cyg). This star possibly is a non-AGB star as it is not a long period, Mira-like variable.

The strong Rb overabundances observed coupled with the lack of strong Zr enhancements in these stars (8) are certainly not predicted by current theoretical models, which usually do not consider stars in this very high mass range, neither take into account the very strong mass loss rates that these stars experience during the TP-AGB phase. The new results here presented are thus challenging our understanding of the late stages of the evolution of intermediate-mass stars.

Remarkably, Rb was found to be not overabundant in the few O-rich massive AGB stars previously studied in the Magellanic Clouds (6), contrary to the Galactic O-rich AGB stars here studied. The luminous O-rich AGBs of the Magellanic Clouds, in fact, may constitute a sample of stars less massive than the present one, in which the temperature for HBB at the bottom of the convective layers is reached more easily than in the Galactic population of AGB stars here considered. Due to their lower metallicity (22), they would develope smaller temperatures during the helium thermal pulse, and thus activate only the $^{13}C(\alpha,n)^{16}O$ neutron chain. The dependence of Rb production and the efficiency of the $^{22}Ne$ neutron source on parameters such as metallicity, mass loss rate, etc. should be fully investigated in the future.

Finally, it is also worthwhile to mention the implications of our work in Meteoritics. If the radioactive chronometer $^{87}Rb$ is strongly overproduced by the activation of the $^{22}Ne$ neutron source in HBB AGB stars, the overall nucleosynthesis of this isotope in the Galaxy may be significantly affected. Huge amounts of Rb-rich processed material can be transferred to the ISM by massive AGB stars with relevant implications for the Rb primeval solar nebula abundance. Radioactive dating studies of primitive chondrites assume that the initial conditions are known and that the oldest components of chondrites (CAIs) evolved without external exchange of $^{87}$Rb and $^{87}$Sr, but our data suggest that the initial $^{87}$Rb/$^{87}$Sr ratio may have been altered by a nearby population of massive AGB stars during the early evolution of our Solar System. The presence of these stars in the vicinity of the Sun is also supported by the identification until date of a few oxide presolar grains for which an intermediate-mass AGB origin has been determined (23, 24). In such environment, it is very unlikely that a Rb/Sr chondritic ratio could be maintained constant in the protoplanetary disk for the time span in which chondritic meteorites formed as previously suggested (25). Consequently, the $^{87}$Rb/$^{87}$Sr ratio measured in primitive meteorites should be considered only a qualitative measure of antiquity, such as was previously pointed out from studies of refractory inclusions in Allende meteorite (26)

27. AM and PGL acknowledge support from grant AYA 2004-3136 and AYA 2003-9499 from the Spanish Ministerio de Educación y Ciencia (MEC). JMTR thanks MEC for a JdC grant. This work is based on observations obtained at the 4.2m WHT telescope operated on the island of La Palma by the Isaac Newton Group in the Spanish Observatorio del Roque de Los Muchachos of the Instituto de Astrofisica de Canarias. Also, based on observations obtained with the 3.6m telescope at ESO-La Silla Observatory (Chile)

**Supplementary Online Material.**

We present here the velocity corrected spectrum of the 60 massive O-rich AGB stars analysed in this work in the region covering the Rb I resonance line at 7800.3 Å, which is marked with a dashed vertical line. The spectra have been normalized and the flux units are arbitrary.

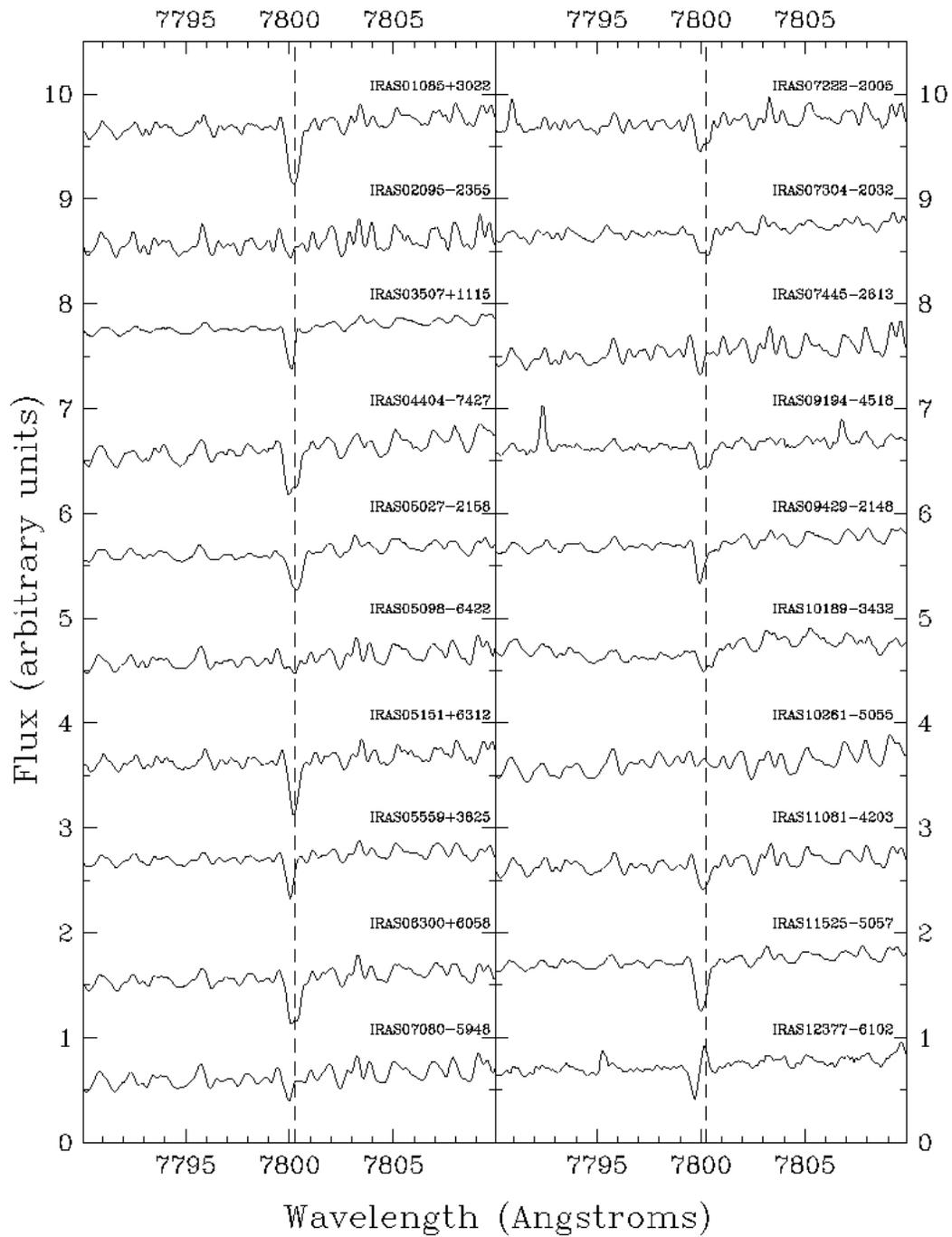

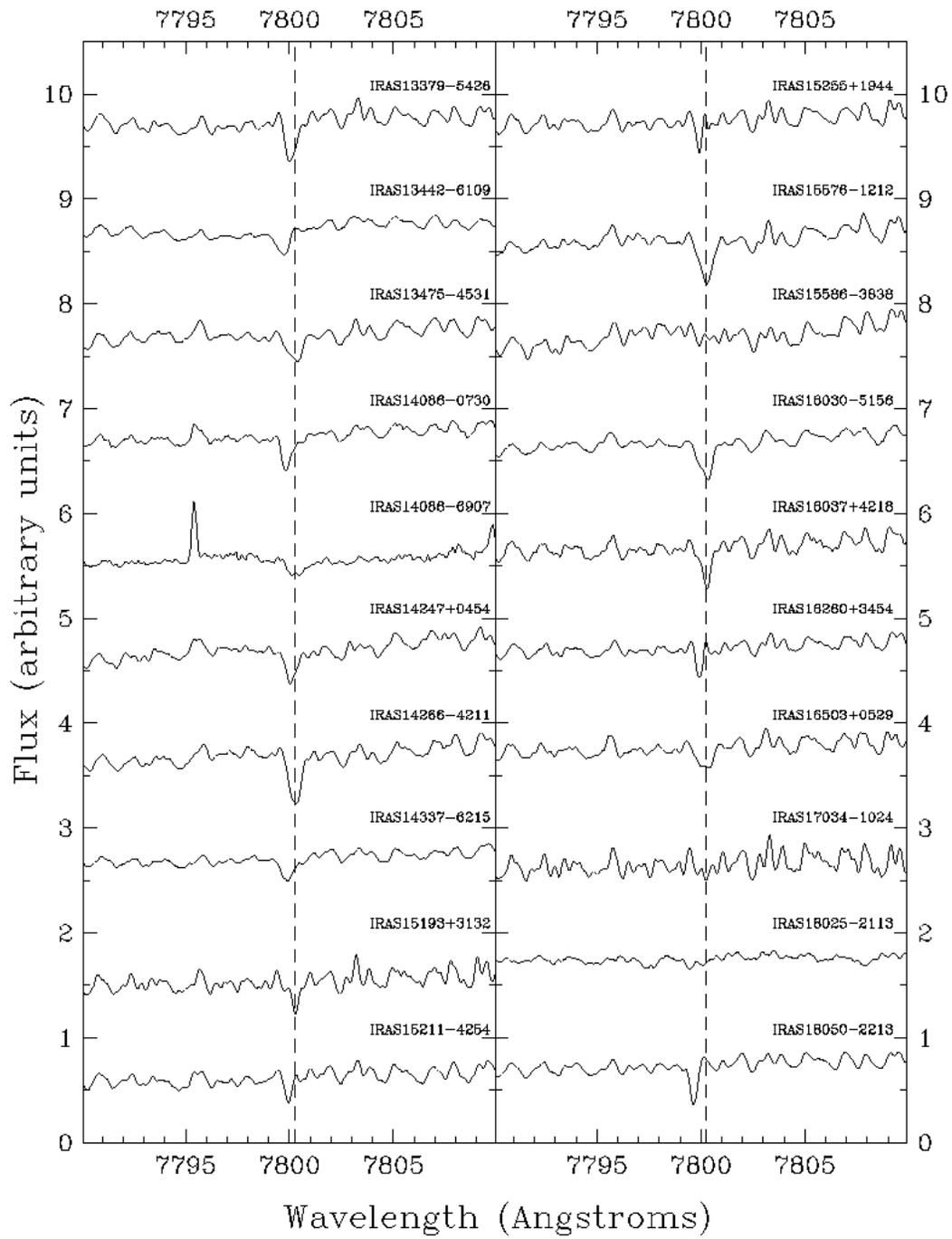

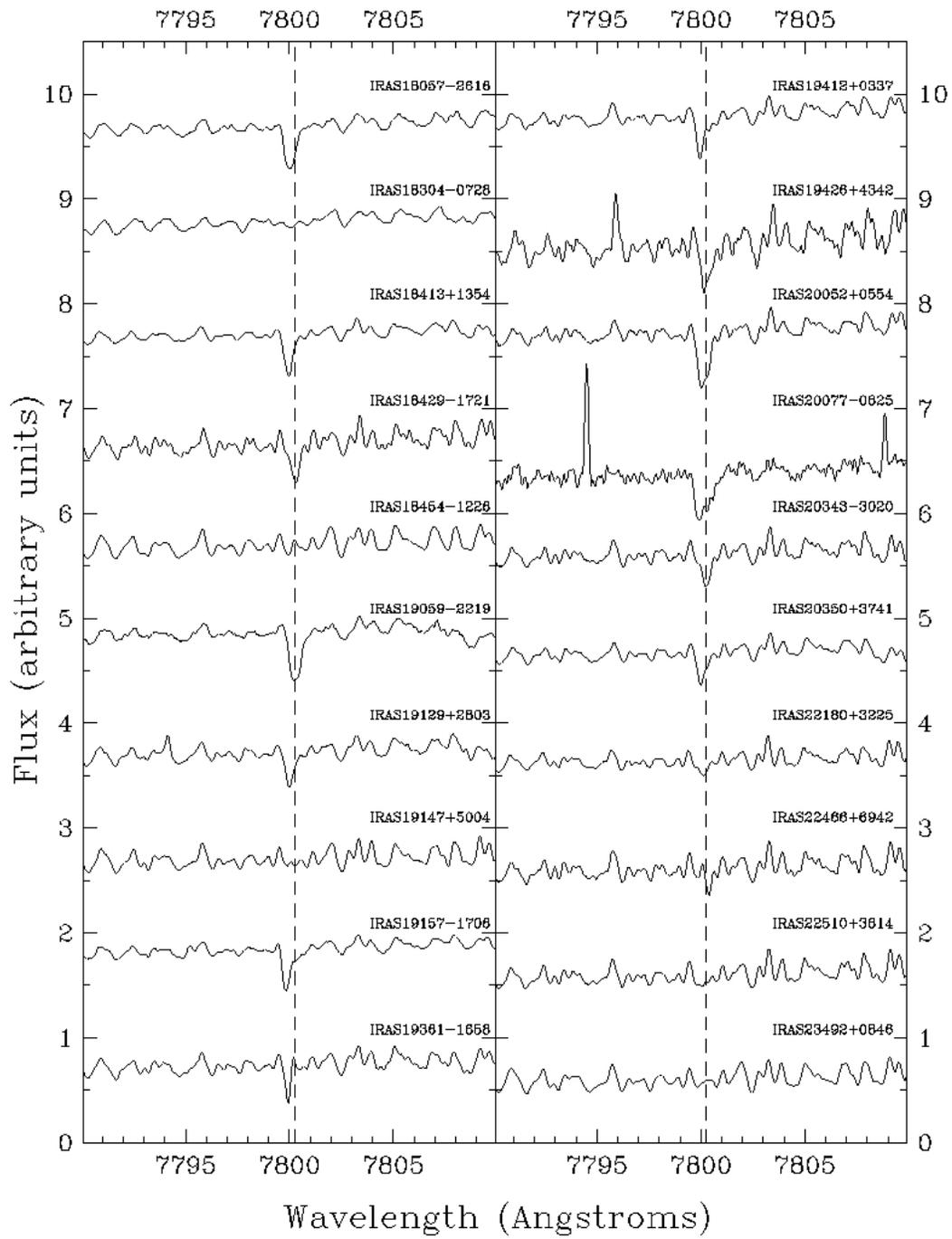